\documentclass{optica-article}

\journal{opticajournal} 

\articletype{Research Article}

\usepackage{siunitx}

\begin{document}

\title{Simultaneous ponderomotive squeezing of light by two mechanical modes in an optomechanical system}

\author{Peyman Malekzadeh\authormark{1}, Emil Zeuthen\authormark{1},Eric Langman\authormark{2}, Albert Schliesser\authormark{2}, Eugene Polzik\authormark{1,*} }

\address{\authormark{1}Niels Bohr Institute, University of Copenhagen, Blegdamsvej 17, 2100 Copenhagen, Denmark}
\address{\authormark{2}Center for Hybrid Quantum Networks, Niels Bohr Institute, University of Copenhagen, Blegdamsvej 17, 2100 Copenhagen, Denmark}
\email{\authormark{*}polzik@nbi.ku.dk} 


\begin{abstract*} 
We experimentally demonstrate a source of squeezed light featuring simultaneous ponderomotive squeezing from two mechanical modes of an optomechanical system. We use ultra-coherent vibrational modes ($Q$ factors on the order of $10^{8}$) of a soft-clamped membrane placed in a Fabry-Pérot optical cavity at cryogenic conditions ($T=11\,\mathrm{K}$) and driven by quantum fluctuations in the intensity of light to create correlations between amplitude and phase quadratures of the intra-cavity light field. Continuous optical monitoring was conducted on two different mechanical modes with a frequency separation of around $1\,\mathrm{MHz}$. As a result of the interaction between the membrane position and the light, we generated ponderomotive squeezing of $4.8\,\mathrm{dB}$ for the first localized mechanical mode at $1.32\,\mathrm{MHz}$ and $4.2\,\mathrm{dB}$ for the second localized mode at $2.43\,\mathrm{MHz}$, as observed in direct detection when correcting for the detection inefficiency. 
Thus, we have demonstrated how squeezed light generation can be extended beyond a single octave in an optomechanical system by leveraging more than one mechanical mode. 
Utilizing homodyne detection to detect squeezing in an optimal quadrature would lead to squeezing levels at the output of the cavity of 7.3\,dB and 6.8\,dB, in the two modes respectively.
Squeezing of light demonstrated here for near-infrared light can be achieved in a broad range of wavelengths due to the relative insensitivity of optomechanical interaction with SiN membranes to the wavelength.  

\end{abstract*}

\section{Introduction}
Quantum mechanics limits the precision with which the state of a light field can be prepared, as characterized, e.g., by the uncertainties in its amplitude and phase quadratures. 
The product of variances of these two conjugate variables has a lower limit, dictated by the Heisenberg Uncertainty Principle, referred to in optical physics as the shot-noise limit \cite{Xiao1987PrecisionLimit}.
Coherent states of light are shot-noise-limited with equal uncertainties in amplitude and phase corresponding to those of the vacuum.  
Squeezing the state of light allows the breaking of this symmetry by reducing the fluctuations in one quadrature below that of the vacuum state while increasing the uncertainty in the conjugate quadrature (``anti-squeezing'').

Following an initial trial involving four-wave mixing that demonstrated the generation of a squeezed state of light \cite{Slusher1985ObservationCavity}, such states have found applications in diverse quantum measurement settings such as gravitational wave detection in LIGO and VIRGO \cite{Tse2019Quantum-EnhancedAstronomy}, mechanical force detection \cite{Mason2019ContinuousLimit}, quantum computing \cite{Zhong2021Phase-ProgrammableLight, Madsen2022QuantumProcessor}, free-electron quantum optics \cite{DiGiulio2020Free-electronLight}, Raman spectroscopy \cite{deAndrade2020Quantum-enhancedSpectroscopy}, back-action-evading measurement \cite{Mller2017QuantumFrame, Yap2020BroadbandInjection}, and quantum sensing \cite{Yang2023ASensing}, among others.
Various physical systems for producing squeezed light have been proposed and implemented, including non-linear optical media \cite{Park2024ParkOscillator, Vahlbruch2016DetectionEfficiency}, atomic ensembles \cite{Jia2023AcousticRegime}, and optomechanical systems \cite{Safavi-Naeini2013SqueezedResonator, Brooks2012Non-classicalOptomechanics, Purdy2014StrongLight}. 
The latter platform has created new opportunities for producing squeezed light by leveraging their capacity to modify the quantum characteristics of light through interactions with mechanical motion, referred to as ponderomotive squeezing. Optomechanical systems encompass diverse configurations and scales, and have demonstrated squeezing in on-chip optomechanical systems \cite{Safavi-Naeini2013SqueezedResonator}, atomic clouds confined within optical cavities \cite{Brooks2012Non-classicalOptomechanics}, and membrane-in-the-middle (MIM) optical cavities \cite{Purdy2014StrongLight, Nielsen2017MultimodeRegime}. 
The MIM systems have achieved a higher level of squeezing due to increased interaction between light and the macroscopic 
mechanical component. Notable recent demonstrations involving a single mechanical mode include, e.g., Refs.~\cite{Mason2019ContinuousLimit,Thomas2021EntanglementSystems}. 
In optomechanical systems with multiple quantum-regime mechanical modes, the highest reported squeezing level is $3.6\,\mathrm{dB}$, demonstrated in a MIM system with a single phononic band gap in the membrane frame \cite{Nielsen2017MultimodeRegime}. In contrast, our demonstration involves a soft-clamped membrane, resulting in much higher $Q$-factors compared to Ref.~\cite{Nielsen2017MultimodeRegime}, and two mechanical modes within \emph{separate} 100-kHz-wide band gaps in the membrane resonator itself. Through the optomechanical interaction, these mechanical resonances, at $1.32\,\mathrm{MHz}$ and $2.43\,\mathrm{MHz}$, lead to the reduction of quantum noise of the output light below the shot-noise level by $4.8\,\mathrm{dB}$ for sideband frequencies near the first mechanical mode and $4.2\,\mathrm{dB}$ near the second mechanical mode, being operated at an ambient temperature of $T=11\,\mathrm{K}$. (The squeezing levels cited here refer to experimentally observed values corrected for detector inefficiency.)

\section{Soft-clamped membrane-in-the-middle optomechanical system}\label{sec:exp-setup}
\begin{figure}[b]
\centering
\includegraphics[width=13cm]{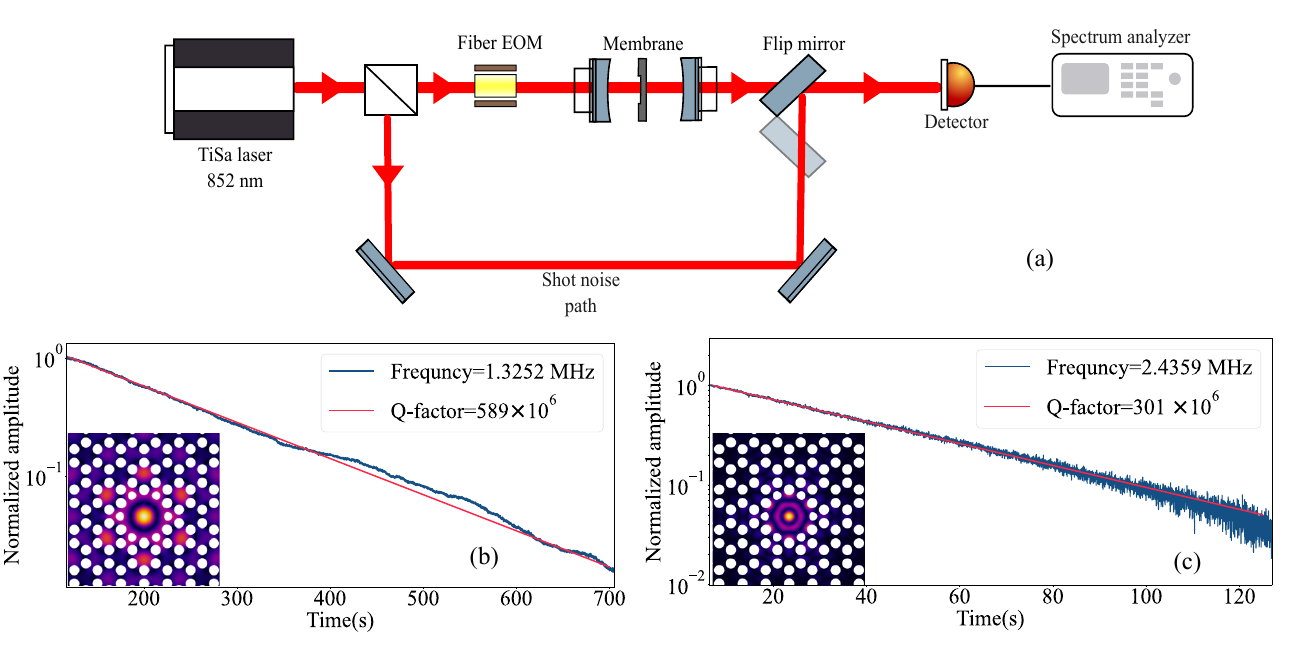}
\caption{(a) Schematic of the experimental setup. The fiber electro-optic modulator (EOM) modulates the input light phase for cavity detuning measurement. The laser shot noise is measured via a separate path and is directed by the flip mirror to the same detector as the squeezing measurement detector. (b,c) Ring-down measurements to extract the $Q$ factor for the first (b) and second (c) localized membrane modes. The embedded pictures depict a sketch of the membrane structure and a simulation of the out-of-plane displacement around the central defect.}
\label{fig1}
\end{figure}
In this research, we employed a Fabry-Pérot optical cavity with two curved mirrors, resulting in a 2.1-mm-long cavity that is overcoupled at $\eta_{\mathrm{cav}0}\equiv\kappa_2/\kappa= 94\%$ (i.e., the angular loss rate from the cavity's transmission port over the cavity linewidth). By placing the membrane inside the cavity, this number decreases to $\eta_{\mathrm{cav}}=91\%$ due to external losses originating from membrane reflection and birefringence. In Fig.~\ref{fig1}(a), a schematic of the detection setup is depicted, where we couple light from the high reflective cavity port and use the transmitted light to lock the cavity and detect the squeezed light. 
The linewidth of the cavity without the membrane was measured to be $\kappa_0/2\pi=5.45\,\mathrm{MHz}$. 
A silicon nitride ($\text{Si}_3 \text{N}_4$) membrane with a thickness of
$20\,\mathrm{nm}$ was employed as the movable mechanical element, featuring an incorporated phononic crystal pattern to create two band gaps and dampen all vibrational modes inside the frequency-band-gap region. 
The mechanical resonant frequency $\Omega_{\mathrm{m}l}$ of the localized mode within each band gap ($l=1,2$) is concentrated near a deliberately created crystal defect, and altering the structure of this defect modifies the mechanical frequency of the dominant vibrational mode inside the defect \cite{Tsaturyan2017UltracoherentDilution}.
By optimizing the cavity assembly's alignment, we ensure that the intracavity field is focused on the membrane's central defect. By moving the cavity mirrors using piezo actuators, we can change the membrane position relative to the cavity standing wave to achieve the maximum optomechanical interaction based on the transfer-matrix model \cite{Jayich2008DispersiveCavity} and, correspondingly, maximum squeezing of quantum noise. 

As depicted in the schematic [Fig.~\ref{fig1}(a)], we employ a direct-detection technique to measure the squeezing of light. Although this method does not permit us to select the quadrature with the greatest squeezing level, we used this approach for experimental simplicity. In the direct-detection method, the light carrier at $\omega_\mathrm{laser}$ serves as a local oscillator for the mechanical sidebands at $\omega_\mathrm{laser}\pm\Omega_{{\mathrm{m}}l}$, analogous to a homodyne detection scheme, but without direct control over the local oscillator phase.

The strong spatial confinement of the mechanical modes, along with complete isolation of the defect from the substrate and reduction in the resonator material's intrinsic dissipation, result in a low dissipation rate in these membranes and a correspondingly high $Q$ factor. At a cryogenic temperature of $T=11\,\mathrm{K}$, the first mechanical mode was measured to have a resonant frequency of $\Omega_{\mathrm{m}1}/2\pi=1.32\,\mathrm{MHz}$ and a ring-down measurement yielded a value of $Q_{\mathrm{m}1}= \Omega_{\mathrm{m}1}/\Gamma_{\mathrm{m}1}=5.89\times10^8$, i.e., an intrinsic damping rate $\Gamma_{\mathrm{m}1}\slash\*2\pi=2.3\,\mathrm{mHz}$ [Fig.~\ref{fig1}(b)]. 
Similarly, the second mechanical mode had a frequency of $\Omega_{\mathrm{m}2}/2\pi=2.43\,\mathrm{MHz}$ and an associated quality factor $Q_{\mathrm{m}2}=3.01\times10^8$, i.e., $\Gamma_{\mathrm{m}2}\slash2\pi=8.1\,\mathrm{mHz}$ [Fig.~\ref{fig1}(c)].
 
Reaching the quantum regime in an optomechanical system involves overcoming thermal decoherence by establishing conditions where the dominant force driving the membrane modes is the intensity fluctuation of light—or, in other words, quantum back action (QBA) caused by the measurement. The ratio of QBA force $S_\mathrm{FF}$ to thermal noise $S_\mathrm{FF}^\mathrm{th}$ spectral density is often characterized via the quantum cooperativity $C_\mathrm{q}$ \cite{Aspelmeyer2014CavityOptomechanics}. 
The primary difficulty in producing ponderomotive squeezing lies in achieving a value of $C_\mathrm{q}\gtrsim 1$. 
In our system we achieve the large-$C_\mathrm{q}$ regime by virtue of three mechanisms: Firstly, by reducing thermal noise through the development of the membrane as discussed in the present section and in Ref.~\cite{Tsaturyan2017UltracoherentDilution}. Secondly, the cavity assembly is suspended in a cryostat, and the system is cooled to $T=11\,\mathrm{K}$. At this temperature, the thermal occupation of the mechanical oscillator modes can be approximated as $n_{\mathrm{th},l}\approx k_\mathrm{B}T/({\hbar\Omega_{\mathrm{m}l}})$ in terms of the Boltzmann constant $k_\mathrm{B}$, resulting in $n_{\mathrm{th},1}\approx\num[group-separator={,}]{172000}$ and $n_{\mathrm{th,2}}\approx\num[group-separator={,}]{94000}$. Thirdly, by increasing the optical drive power, thereby increasing the QBA force.

\section{Theory}\label{sec:theory}
A general quadrature of the light leaving the optical cavity $\hat{X}_{\mathrm{out}}^{(\theta)}\equiv\hat{X}_{\mathrm{out}}\cos\theta+\hat{Y}_{\mathrm{out}}\sin\theta$ is a linear combination of the amplitude $\hat{X}_{\mathrm{out}}$ and phase $\hat{Y}_{\mathrm{out}}$ quadratures. A quadrature is said to be squeezed whenever its fluctuations, as measured by its power spectral density, are less than those of a vacuum state $S[X_{\mathrm{out}}^{(\theta)}]<S_\mathrm{SN}$. 
This is accompanied by an increase in the fluctuations of the orthogonal quadrature $\hat{X}_{\mathrm{out}}^{(\theta+\pi/2)}$, $S[X_{\mathrm{out}}^{(\theta+\pi/2)}]>S_\mathrm{SN}$, i.e., anti-squeezing.
In the direct-detection scheme employed here (as described in the previous section), we are constrained to measuring the amplitude quadrature $\hat{X}_{\mathrm{out}}$ (see also Supplementary Material Subsec.~1.1).

\subsection{The ponderomotive squeezing mechanism}\label{subsec:squeezing-mechanism}
We will now describe how the optomechanical interaction leads to squeezed fluctuations in certain quadratures.
The Hamiltonian describing the radiation-pressure interaction between the intracavity light mode and mechanical oscillator mode $l$ is 
\begin{equation}\label{eq:H-int}
\hat{H}_{\mathrm{int},l}/\hbar = 2g_l \hat{X} \hat{Q}_l = g_l(\hat{a}+\hat{a}^\dagger)(\hat{b}_l+\hat{b}_l^{\dagger}),
\end{equation}
where the ladder operators of the optical mode $[\hat{a},\hat{a}^\dagger]=1$ (in the rotating frame of the laser drive $\omega_\mathrm{laser}$) and the mechanical mode $l$, $[\hat{b}_l,\hat{b}_{l'}^\dagger]=\delta_{l,l'}$, have been introduced via the dimensionless optical amplitude quadrature $\hat{X}=(\hat{a}+\hat{a}^\dagger)/\sqrt{2}$ and the dimensionless mechanical position $\hat{Q}_l=(\hat{b}_l+\hat{b}_l^\dagger)/\sqrt{2}$; their canonical conjugates are the phase quadrature $\hat{Y}=-i(\hat{a}-\hat{a}^\dagger)/\sqrt{2}$ and the dimensionless mechanical momentum $\hat{P}_l=-i(\hat{b}_l-\hat{b}_l^\dagger)/\sqrt{2}$, respectively. 
The drive-enhanced optomechanical interaction rate can be written as $g_{l}=\sqrt{n_{\mathrm{ph}}} x_{\mathrm{zpf},l}G_{l}$, where $G_{l}$ represents the optical frequency change per displacement of the mechanical oscillator, $x_{\mathrm{zpf},l}=\sqrt{\hbar\slash(\*2m_{l}\Omega_{\mathrm{m}l})}$ is the zero-point fluctuation amplitude, $m_{l}$ is the oscillator effective mass, and $n_\mathrm{ph}$ signifies the mean number of drive photons contained within the cavity. 
Equation~\eqref{eq:H-int} can be decomposed into its beam-splitter $g (\hat{a}\hat{b}_l^\dagger+\hat{a}^\dagger\hat{b}_l)$ and two-mode-squeezing $g (\hat{a}\hat{b}_l+\hat{a}^\dagger\hat{b}_l^\dagger)$ components; 
it is a well-known feature of optomechanics that their effective strength can be adjusted via the detuning $\Delta=\omega_\mathrm{cav}-\omega_\mathrm{laser}$ of the cavity resonance from the frequency of the coherent laser drive, seeing as this determines to which extent they are energy conserving, in an interplay with the cavity linewidth $\kappa$ and the mechanical resonance frequency $\Omega_{\mathrm{m},l}$ \cite{Aspelmeyer2014CavityOptomechanics}. Working with a red-detuned drive $\Delta>0$ favors the stabilizing optical cooling dynamics of the beam-splitter interaction, permitting continuous, steady-state operation. 

The Heisenberg equation of motion and Eq.~\eqref{eq:H-int} imply that the mechanical momentum $\hat{P}_l$ is subjected to a QBA force $\propto \hat{X}$, whose influence, in turn, rotates into the mechanical position $\hat{Q}_l$ due to the free mechanical evolution, as governed by $\hat{H}_{\mathrm{m},l}/\hbar=\Omega_{\mathrm{m}l}(\hat{Q}_l^2+\hat{P}_l^2)/2$.
On account of the Heisenberg equation of motion and Eq.~\eqref{eq:H-int} yet again, the QBA signal maps from $\hat{Q}_l$ into the phase quadrature of the intracavity light $\hat{Y}$. This is the essence of the ponderomotive squeezing mechanism that induces correlations between the amplitude and phase quadratures of light. 

To witness these correlations generally requires the detection of a rotated quadrature of the output light $\hat{X}_{\mathrm{out}}^{(\theta)}$. However, the free evolution of the cavity mode, governed by
\begin{equation}\label{eq:H-cav}
    \hat{H}_\mathrm{cav}/\hbar = \frac{\Delta}{2}(\hat{X}^2 + \hat{Y}^2) = \Delta (\hat{a}^\dagger \hat{a}+1/2),
\end{equation}
is in itself a rotation of the optical quadratures in the $(\hat{X},\hat{Y})$ plane at rate $\Delta$. This entails that even if we detect the amplitude quadrature of the outgoing light at the detection port (labelled `2') $\hat{X}_{\mathrm{out},2}=\sqrt{\kappa_2}\hat{X}-\hat{X}_{\mathrm{in},2}$, the interference between the intracavity $\propto \hat{X}$ and reflected $\hat{X}_{\mathrm{in},2}$ components implies that squeezing may be observed in the direct-detection scheme considered here, insofar as $\Delta\neq 0$.

\subsection{Squeezing spectrum}
Above we described qualitatively how the coupling between optical and mechanical subsystems causes the input optical fluctuations to interfere in the optical field leaving the cavity. 
To demonstrate this mechanism quantitatively and derive the expected form of the observed spectrum of light, we proceed to solving the Heisenberg-Langevin equations of the optomechanical system in the Fourier-frequency domain. This allows us to obtain the position quadrature for mechanical modes $l=1,2$ [see Supplementary Material Sec.~1 for details],
\begin{equation}
    {\hat{Q}_l(\omega)=\chi_{\mathrm{eff}l}}(\omega)\sqrt{2\Gamma_{\mathrm{m}l}}\hat{P}_{\mathrm{in},l}(\omega)-\frac{2g_l\chi_{\mathrm{eff}l}(\omega)}{(\kappa/2-i\omega)^2+\Delta^2}\sum_{j}\sqrt{\kappa_{j}}\left[(\kappa/2-i\omega)\hat{X}_{\mathrm{in},j}(\omega)+\Delta\hat{Y}_{\mathrm{in},j}(\omega)\right],
    \label{equation:3}
\end{equation}
where $\hat{X}_{\mathrm{in},j}$  
and $\hat{Y}_{\mathrm{in},j}$ 
represent the amplitude and phase fluctuations of the incoming light, which are imprinted on the membrane motion from each of the cavity ports $j=1$ (input), 2 (output), and ext (external losses), weighted according to their respective decay rates $\kappa_j$ (for which $\sum_j\kappa_j=\kappa$). 
Additionally, the $\hat{P}_{\mathrm{in},l}$  
operator represents the mechanical momentum input fluctuations associated with the damping $\Gamma_{\mathrm{m}l}$. 
In Eq.~\eqref{equation:3}, we have introduced the effective mechanical susceptibility, which governs the mechanical system's response to applied forces $\chi_{\mathrm{eff}l}(\omega)=(\chi_{\mathrm{m}l}^{-1}(\omega)
-4g_l^2\Delta/[(\kappa/2-i{\omega})^2+\Delta^2])^{-1}$, defined in terms of the bare susceptibility $\chi_{\mathrm{m}l}(\omega)=\Omega_{\mathrm{m}l}/(\Omega_{\mathrm{m}l}^2-\omega^2-i\Gamma_{\mathrm{m}l}\omega)$. 
The effective mechanical susceptibility exhibits a shifted resonance frequency and modified linewidth due to the dynamical back-action loop; this is another consequence of the cavity quadrature rotation occurring when $\Delta\neq 0$. 
Crucially, in deriving Eq.~\eqref{equation:3} 
we have assumed that the effective susceptibilities of the two mechanical modes are essentially non-overlapping $\chi_{\mathrm{eff}1}(\omega)\chi_{\mathrm{eff}2}(\omega)\approx 0$, as is the case in the present experiment.

Turning now to the response of the amplitude quadrature of the intracavity light, we find it to be given by
\begin{equation}
\hat{X}(\omega)=\frac{1}{(\kappa/2-i\omega)^2+\Delta^2}\left(\sum_{j}\sqrt{\kappa_{j}}\left[(\kappa/2-i\omega)\hat{X}_{\mathrm{in},j}(\omega)+\Delta\hat{Y}_{\mathrm{in},j}(\omega)\right]-2\Delta\sum_l g_l\hat{Q}_l(\omega)\right),
\label{equation:4}
\end{equation}
showing that the mapping of the mechanical response into the amplitude quadrature $\hat{X}$ hinges on the quadrature rotation afforded by $\Delta\neq 0$, as discussed in the context of Eq.~\eqref{eq:H-cav}.
By substituting Eq.~(\ref{equation:3}) into Eq.~(\ref{equation:4}) and employing the input-output relation $\hat{X}_{\mathrm{out},j}=\sqrt{\kappa_j}\hat{X}-\hat{X}_{\mathrm{in},j}$, we can obtain an expression for the output port amplitude quadrature $\hat{X}_{\mathrm{out},2}$ in terms of the input operators $\hat{X}_{\mathrm{in},j}$, $\hat{Y}_{\mathrm{in},j}$, and $\hat{P}_{\mathrm{in},l}$ (where $j=1,2,\mathrm{ext}$ and $l=1,2$).
The corresponding power spectral density for frequencies in the vicinity of the resonance of mechanical mode $l$, $\omega\sim\Omega_{\mathrm{m}l}$, is 
\begin{multline}\label{equation:5}
    S_{X_{\mathrm{out},2}}(\omega\sim \Omega_{\mathrm{m}l}) = 
    \eta_{\mathrm{cav}}\left|\sqrt{\kappa}\frac{2\Delta g_l\chi_{\mathrm{eff}l}(\omega)\sqrt{2\Gamma_{\mathrm{m}l}}}{\left(\kappa/2 - i\omega\right)^2 + \Delta^2}\right|^2 
    \left(n_{\mathrm{th},l} + \frac{1}{2}\right) \\
    +\eta_{\mathrm{cav}} \left|\frac{\kappa\chi_{\mathrm{eff}l}(\omega)/\chi_{\mathrm{m}l}(\omega)}{\left(\kappa/2 - i\omega\right)^2 + \Delta^2} 
    \right|^2 
    \left[\frac{\kappa^2}{4} + \omega^2+ \Delta^2 \right]\frac{1}{2} 
    -\eta_{\mathrm{cav}}\text{Re}\left(\frac{(\kappa/2 - i\omega)\kappa}{\left(\kappa/2 - i\omega\right)^2 + \Delta^2}\frac{\chi_{\mathrm{eff}l}(\omega)}{\chi_{\mathrm{m}l}(\omega)}\right)
    + \frac{1}{2},
\end{multline}
where we have used $\eta_\mathrm{cav}=\kappa_2/\kappa=1-(\kappa_1+\kappa_\mathrm{ext})/\kappa$ and, again, the assumption that the effective susceptibilities of the two mechanical modes are essentially non-overlapping $\chi_{\mathrm{eff},1}(\omega)\chi^*_{\mathrm{eff},2}(\omega)\approx 0$.
Finally, we account for the finite detection efficiency $\eta_\mathrm{det}$, which entails the admixture of uncorrelated vacuum noise according to
\begin{equation}\label{eq:S-det}
S_\mathrm{det}(\omega)= \eta_\mathrm{det} S_{X_{\mathrm{out},2}}(\omega)+(1-\eta_\mathrm{det})\frac{1}{2}.
\end{equation}
Note that the shot-noise level equals $1/2$ in the convention employed in Eqs.~\eqref{equation:5} and \eqref{eq:S-det}. 
Equation~\eqref{eq:S-det} is the final result of this section, and below we fit it to the experimental spectra in the vicinities of the two mechanical modes $\omega\sim\Omega_{\mathrm{m}1},\Omega_{\mathrm{m}2}$.

\section{Results}\label{sec:results}
\begin{figure}[bht]
\centering\includegraphics[width=13.5cm]{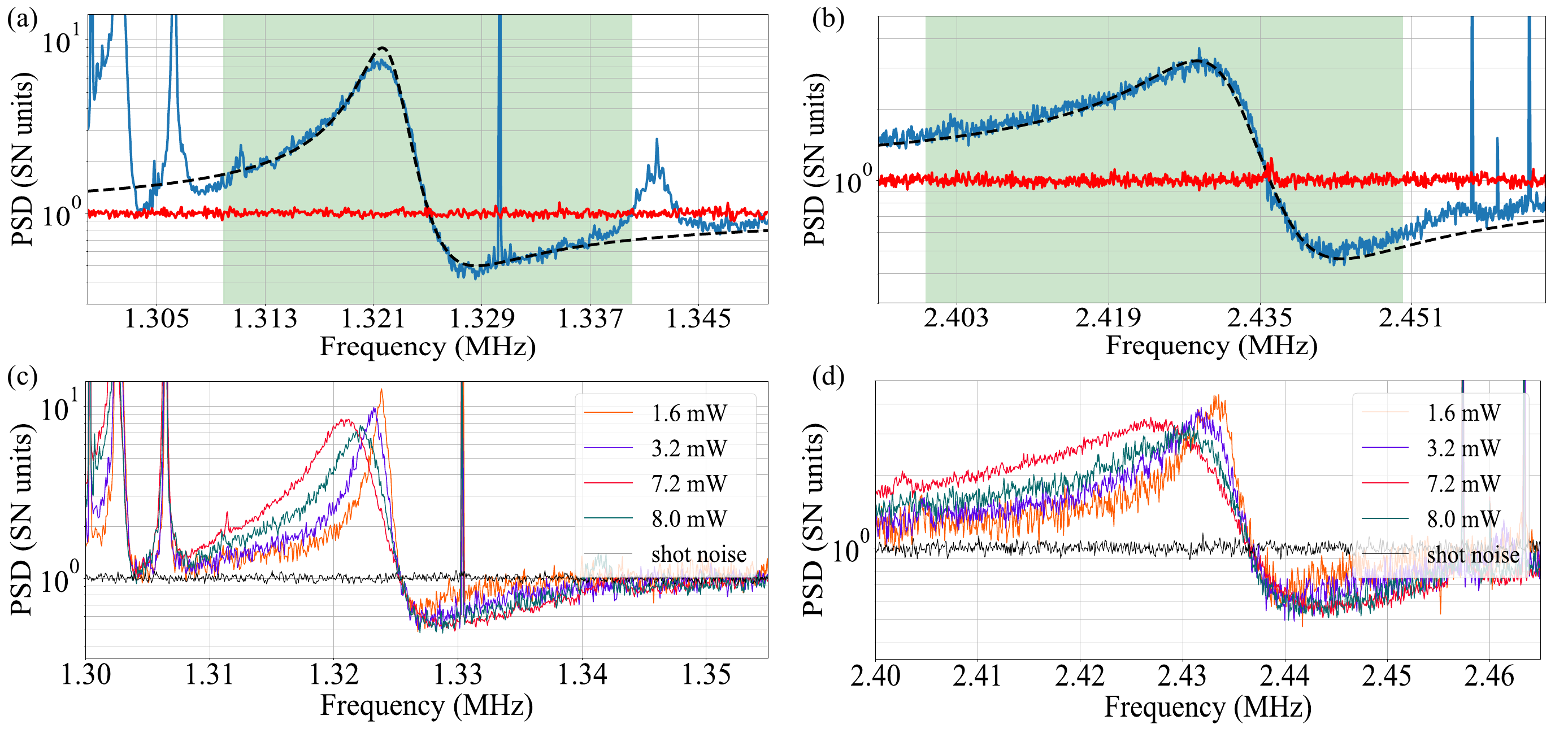}
\caption{Spectra of the light output from the optomechanical system for $7.1\,\mathrm{mW}$ input light coupled to the cavity and measured by direct detection. (a,b) Spectrum exhibiting squeezing near the resonance of the (a) first and (b) second localized mechanical mode. Blue traces are cavity output spectra, shot noise is indicated in red, the dashed black line is the model fitted on the spectra for each mode and the green areas specify the fitting region. The laser classical noise around $1.33\,\mathrm{MHz}$ in (a) and out-of-band-gap modes are not included in the fitting model. (c,d) The output squeezed spectrum is recorded at different input power to the cavity levels, from 1.6\,mW to 8\,mW, in the first (c) and second (d) membrane band gaps.}
\label{fig:2}
\end{figure}
We measured the power spectral density near the two mechanical modes by directly observing the light transmitted from the cavity, as shown in Fig.~\ref{fig:2}. 
The vacuum light state is measured using a dedicated pathway from the TiSa laser to the same detector [Fig.~\ref{fig1}(a)]. The electronic noise in the detection pathway is subtracted from the experimental results and shot-noise traces. The optical path and detection efficiency is $\eta_\mathrm{det}= 83\%$. 

We fit the theoretical model~\eqref{eq:S-det} to the experimental data by minimizing the sum of squares of their relative deviations, using as fit parameters the cavity linewidth $\kappa$, the drive detuning $\Delta$, the optomechanical couplings $g_l$, and mechanical resonant frequencies $\Omega_{\mathrm{m}l}$ ($l=1,2$), whereas the values for $\Gamma_{\mathrm{m}l}$, and $n_{\mathrm{th},l}$ as well as $\eta_\mathrm{cav}$ and $\eta_\mathrm{det}$ were established via separate calibration measurements as described in Sec.~\ref{sec:exp-setup}. 
The fitted detuning value was $\Delta/2\pi=2.3\,\mathrm{MHz}$ and the fitted cavity linewidth (FWHM) value was $\kappa/2\pi=7.0\,\mathrm{MHz}$. The fitted values for $\Omega_{\mathrm{m}l}$ were consistent with the ring-down measurements (see Sec.~\ref{sec:exp-setup}).
From the fitted values of $g_l$ we can calculate for each mechanical mode the measurement rate $\Gamma_{\mathrm{meas},l}=4g_l^2/\kappa$, characterizing the transfer of information into the outgoing light fields (insofar as $g_l\ll\kappa$): 
The first mode has a measurement rate of $\Gamma_{\mathrm{meas},1}/2\pi= 7.3\,\mathrm{kHz}$ corresponding to a quantum cooperativity $C_{\mathrm{q}l}=\Gamma_{\mathrm{meas},l}/[\Gamma_{\mathrm{m},l}(n_{\mathrm{th},l}+1/2)]$ equal to $C_{\mathrm{q}1}=3.0$, while the second mode has $\Gamma_{\mathrm{meas},2}/2\pi= 19.0\,\mathrm{kHz}$ and $C_{\mathrm{q}2}=3.9$.

We experimentally observe ponderomotive squeezing of $3.5\,\mathrm{dB}$ and $3.2\,\mathrm{dB}$ below the shot-noise level, near the resonance of the first and second mechanical mode, respectively [Fig.~\ref{fig:2}(a,b)]. The fitted curve is in good agreement with the observed spectra. Correcting the experimental spectra for the detection loss $1-\eta_\mathrm{det}=17\%$ using a relation equivalent to Eq.~\eqref{eq:S-det},  
$S_\mathrm{det}(\omega)= \eta_\mathrm{det} S_{\mathrm{corr}}(\omega)+(1-\eta_\mathrm{det})S_\mathrm{SN}$, 
results in a squeezing level of $4.8\,\mathrm{dB}$ for the first mechanical mode and $4.2\,\mathrm{dB}$ for the mode in the second band gap, which is comparable with the state of the art for optomechanical systems \cite{Safavi-Naeini2013SqueezedResonator,Brooks2012Non-classicalOptomechanics,Purdy2014StrongLight,Nielsen2017MultimodeRegime}. 
All plotted spectra are normalized so that the shot-noise level is unity, $S_\mathrm{SN}=1$. 

\begin{figure}[t]
\centering
\includegraphics[width=1\linewidth]{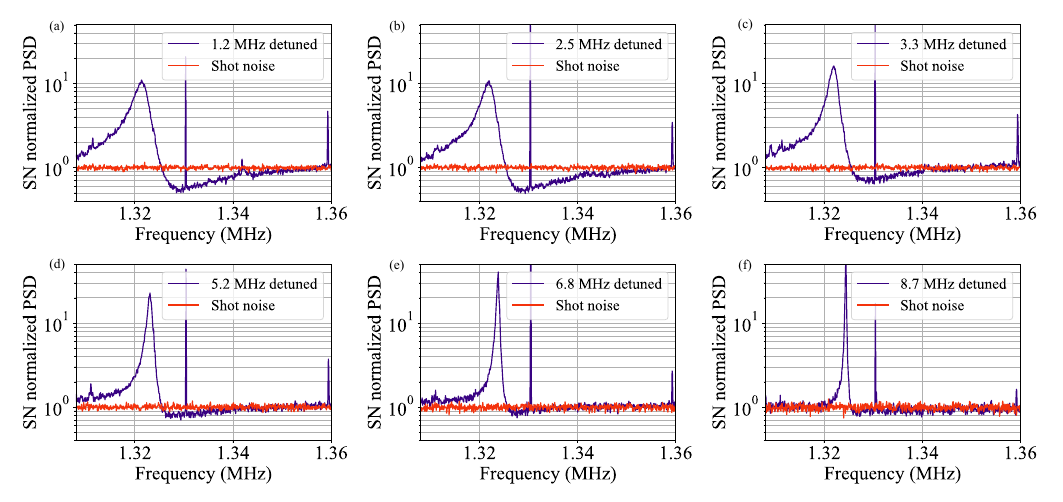}
\caption{Squeezing measurements for the first mechanical mode for different drive detunings from the cavity resonance (spectra for the second mechanical mode are included in Supplementary Material Sec.~2). The cavity input power is $3\,\mathrm{mW}$ for all measurements while detuning (measured via the cavity response to phase modulation of the input light) is swept from $1.2\,\mathrm{MHz}$ to $8.7\,\mathrm{MHz}$.}
\label{fig:3}
\end{figure}

As previously stated, achieving the quantum limit requires the measurement rate to surpass the thermal decoherence rate, $C_{\mathrm{q}l}\gtrsim 1$. The measurement rate is influenced by the intracavity photon number $n_\mathrm{ph}$, which can alter the optomechanical coupling and consequently affect the squeezing levels. Increasing the input power from $1.6\,\mathrm{mW}$ to $8\,\mathrm{mW}$, as depicted in Fig.~\ref{fig:2}(c,d), induced a broadening of the localized mechanical modes. This, in turn, led to an enhanced squeezing level up to an input power of $7.2\,\mathrm{mW}$, but further increasing the input power resulted in a saturation of the squeezing level.
In Fig.~\ref{fig:3}, we varied the laser detuning from the cavity resonance while maintaining a constant input power of $3~\mathrm{mW}$. We recorded the power spectral density of the cavity output for each detuning. The cavity lock point was measured using the cavity response to a phase modulation applied to the input light using an EOM [Fig.~\ref{fig1}a]. The detuning was swept from $1.2\,\mathrm{MHz}$ to $8.75\,\mathrm{MHz}$, revealing an increase in intensity within the cavity closer to resonance and subsequently leading to enhanced optomechanical coupling as anticipated. Analogous to the power sweep shown in Fig.~\ref{fig:2}(c,d), we observe a saturation of the squeezing level at around the detuning of $2.5\,\mathrm{MHz}$ as indicated by the similar squeezing level observed at the detuning of $1.2\,\mathrm{MHz}$. We ascribe this to the complex interplay between the increasing intracavity drive power for decreasing detuning $\Delta$ and the cavity-induced quadrature rotation (at rate $\Delta$) required to observe squeezing in our direct-detection scheme, as discussed in Subsec.~\ref{subsec:squeezing-mechanism}.

\section{Conclusion}
In summary, we have presented a membrane-in-the-middle optomechanical system that squeezes light by coupling it to two distinct mechanical modes at $1.32\,\mathrm{MHz}$ and $2.43\,\mathrm{MHz}$. 
By improving the cavity alignment, enhancing the light-membrane interaction through optimal coupling of the cavity fundamental mode to the membrane defect, and designing the mirror geometry to reduce thermal fluctuations near the mechanical frequencies, as well as engineering the membrane to achieve high quality factors, stably locking the optical cavity, and probing the membrane motion at cryogenic temperatures, 
we were able to realize more than 4\,dB of ponderomotive squeezing (corrected
for detector inefficiency) for both mechanical modes, which is comparable to the state of the art in single-mode optomechanical systems.  

The direct-detection scheme employed here, while experimentally simple, has the drawback that, generally, we cannot observe the quadrature of light that has maximal squeezing (for a given spectral component $\omega$); a full experimental characterization of the generated squeezing would require homodyne detection. Instead we here use the theoretical model and the system parameters to predict the maximal squeezing level that would have been seen in homodyne detection of the optimal light quadrature (see SM Subsec.~1.2). The resulting values for the observed squeezing are 4.9\,dB and 4.7\,dB near the resonance of the first and second mechanical mode, respectively, assuming the same detection efficiency $\eta_{\mathrm{det}}$ as in the present setup. Correcting for the detection inefficiency, we predict the corresponding squeezing levels at the output of the cavity to be 7.3\,dB and 6.8\,dB.

Furthermore, we experimentally investigated the effects of cavity detuning and intracavity power on the squeezing level by sweeping these parameters and measuring the output spectra. By fitting the theoretical model, we extracted important parameters such as the optomechanical coupling and the resulting measurement rate into the output light field, and demonstrated agreement between the experimental data and the theoretical model.

The method described in the present paper allows to generate squeezed light within a very broad range of visible and near-infrared light. Silicon nitride has extremely low absorption losses below 1\,dB/cm in the wavelength range from 400\,nm to 2350\,nm \cite{Blumenthal2018NarrowLasers}. With the 20-nm-thick membrane used in our work, the intracavity losses due to the membrane absorption can be kept below $10^{-6}$ within that wavelength range, paving the way to generation of highly squeezed light for any wavelength of interest in the visible or near-infrared domains.

\begin{backmatter}
\bmsection{Funding}
This work is supported by VILLUM FONDEN under a Villum Investigator Grant, grant no.\ 25880, by the Novo Nordisk Foundation through Copenhagen Center for Biomedical Quantum Sensing, grant number NNF24SA0088433, and by the European Union's Horizon 2020 research and innovation program under the Marie Sklodowska-Curie grant agreement No.\ 847523 `INTERACTIONS'.

\bmsection{Acknowledgment}
 We thank Sergey Fedorov for his contributions to the setup and the experimental data, and Mads Bjerregaard Kristensen for characterizing the membrane at room temperature.
 
\bmsection{Contributions}
P.M.\ ran the experiment and analyzed the experimental data. P.M.\ and E.Z.\ did the theoretical derivations. E.L.\ and A.S.\ provided the membrane, and E.P.\ supervised the project. P.M., E.Z., and E.P.\ wrote the initial draft of the manuscript.

\end{backmatter}

\bibliography{references}

\begin{thebibliography}{10}
\newcommand{\enquote}[1]{``#1''}

\bibitem{Xiao1987PrecisionLimit}
M.~Xiao, L.-A. Wu, and H.~J. Kimble, \enquote{{Precision Measurement beyond the
  Shot-Noise Limit},} {\protect\JournalTitle{Phys. Rev. Lett.}} \textbf{59},
  278--281 (1987).

\bibitem{Slusher1985ObservationCavity}
R.~E. Slusher, L.~W. Hollberg, B.~Yurke, \emph{et~al.}, \enquote{{Observation
  of Squeezed States Generated by Four-Wave Mixing in an Optical Cavity},}
  {\protect\JournalTitle{Phys. Rev. Lett.}} \textbf{55}, 2409--2412 (1985).

\bibitem{Tse2019Quantum-EnhancedAstronomy}
M.~Tse, H.~Yu, N.~Kijbunchoo, \emph{et~al.}, \enquote{{Quantum-Enhanced
  Advanced LIGO Detectors in the Era of Gravitational-Wave Astronomy},}
  {\protect\JournalTitle{Physical Review Letters}} \textbf{123} (2019).

\bibitem{Mason2019ContinuousLimit}
D.~Mason, J.~Chen, M.~Rossi, \emph{et~al.}, \enquote{{Continuous force and
  displacement measurement below the standard quantum limit},}
  {\protect\JournalTitle{Nature Physics}} \textbf{15}, 745--749 (2019).

\bibitem{Zhong2021Phase-ProgrammableLight}
H.-S. Zhong, Y.-H. Deng, J.~Qin, \emph{et~al.}, \enquote{{Phase-Programmable
  Gaussian Boson Sampling Using Stimulated Squeezed Light},}
  {\protect\JournalTitle{Phys. Rev. Lett.}} \textbf{127} (2021).

\bibitem{Madsen2022QuantumProcessor}
L.~S. Madsen, F.~Laudenbach, M.~F. Askarani, \emph{et~al.}, \enquote{{Quantum
  computational advantage with a programmable photonic processor},}
  {\protect\JournalTitle{Nature}} \textbf{606}, 75--81 (2022).

\bibitem{DiGiulio2020Free-electronLight}
V.~Di~Giulio and F.~J. Garc{\'{i}}a~de Abajo, \enquote{{Free-electron shaping
  using quantum light},} {\protect\JournalTitle{Optica}} \textbf{7}, 1820
  (2020).

\bibitem{deAndrade2020Quantum-enhancedSpectroscopy}
R.~B. de~Andrade, H.~Kerdoncuff, K.~Berg-S{\o}rensen, \emph{et~al.},
  \enquote{{Quantum-enhanced continuous-wave stimulated Raman scattering
  spectroscopy},} {\protect\JournalTitle{Optica}} \textbf{7}, 470 (2020).

\bibitem{Mller2017QuantumFrame}
C.~B. M{\o}ller, R.~A. Thomas, G.~Vasilakis, \emph{et~al.}, \enquote{{Quantum
  back-Action-evading measurement of motion in a negative mass reference
  frame},} {\protect\JournalTitle{Nature}} \textbf{547}, 191--195 (2017).

\bibitem{Yap2020BroadbandInjection}
M.~J. Yap, J.~Cripe, G.~L. Mansell, \emph{et~al.}, \enquote{{Broadband
  reduction of quantum radiation pressure noise via squeezed light injection},}
   (2020).

\bibitem{Yang2023ASensing}
W.~Yang, W.~Diao, C.~Cai, \emph{et~al.}, \enquote{{A Bright Squeezed Light
  Source for Quantum Sensing},} {\protect\JournalTitle{Chemosensors}}
  \textbf{11} (2023).

\bibitem{Park2024ParkOscillator}
T.~Park, H.~Stokowski, V.~Ansari, \emph{et~al.}, \enquote{{Park et al
  Single-mode squeezed-light generation and tomography with an integrated
  optical parametric oscillator},} {\protect\JournalTitle{Sci. Adv}}
  \textbf{10}, 1814 (2024).

\bibitem{Vahlbruch2016DetectionEfficiency}
H.~Vahlbruch, M.~Mehmet, K.~Danzmann, and R.~Schnabel, \enquote{{Detection of
  15 dB Squeezed States of Light and their Application for the Absolute
  Calibration of Photoelectric Quantum Efficiency},}
  {\protect\JournalTitle{Physical Review Letters}} \textbf{117} (2016).

\bibitem{Jia2023AcousticRegime}
J.~Jia, V.~Novikov, T.~B. Brasil, \emph{et~al.}, \enquote{{Acoustic frequency
  atomic spin oscillator in the quantum regime},} {\protect\JournalTitle{Nature
  Communications}} \textbf{14} (2023).

\bibitem{Safavi-Naeini2013SqueezedResonator}
A.~H. Safavi-Naeini, S.~Gr{\"{o}}blacher, J.~T. Hill, \emph{et~al.},
  \enquote{{Squeezed light from a silicon micromechanical resonator},}
  {\protect\JournalTitle{Nature}} \textbf{500}, 185--189 (2013).

\bibitem{Brooks2012Non-classicalOptomechanics}
D.~W. Brooks, T.~Botter, S.~Schreppler, \emph{et~al.}, \enquote{{Non-classical
  light generated by quantum-noise-driven cavity optomechanics},}
  {\protect\JournalTitle{Nature}} \textbf{488}, 476--480 (2012).

\bibitem{Purdy2014StrongLight}
T.~P. Purdy, P.~L. Yu, R.~W. Peterson, \emph{et~al.}, \enquote{{Strong
  optomechanical squeezing of light},} {\protect\JournalTitle{Physical Review
  X}} \textbf{3} (2014).

\bibitem{Nielsen2017MultimodeRegime}
W.~H.~P. Nielsen, Y.~Tsaturyan, C.~B. M{\o}ller, \emph{et~al.},
  \enquote{{Multimode optomechanical system in the quantum regime},}
  {\protect\JournalTitle{Proceedings of the National Academy of Sciences of the
  United States of America}} \textbf{114}, 62--66 (2017).

\bibitem{Thomas2021EntanglementSystems}
R.~A. Thomas, M.~Parniak, C.~{\O}stfeldt, \emph{et~al.}, \enquote{{Entanglement
  between distant macroscopic mechanical and spin systems},}
  {\protect\JournalTitle{Nature Physics}} \textbf{17}, 228--233 (2021).

\bibitem{Tsaturyan2017UltracoherentDilution}
Y.~Tsaturyan, A.~Barg, E.~S. Polzik, and A.~Schliesser, \enquote{{Ultracoherent
  nanomechanical resonators via soft clamping and dissipation dilution},}
  {\protect\JournalTitle{Nature Nanotechnology}} \textbf{12}, 776--783 (2017).

\bibitem{Jayich2008DispersiveCavity}
A.~M. Jayich, J.~C. Sankey, B.~M. Zwickl, \emph{et~al.}, \enquote{{Dispersive
  optomechanics: A membrane inside a cavity},} {\protect\JournalTitle{New
  Journal of Physics}} \textbf{10} (2008).

\bibitem{Aspelmeyer2014CavityOptomechanics}
M.~Aspelmeyer, T.~J. Kippenberg, and F.~Marquardt, \enquote{{Cavity
  optomechanics},} {\protect\JournalTitle{Reviews of Modern Physics}}
  \textbf{86}, 1391--1452 (2014).

\bibitem{Blumenthal2018NarrowLasers}
D.~J. Blumenthal, S.~Gundavarapu, G.~M. Brodnik, \emph{et~al.},
  \enquote{{Narrow Linewidth Stimulated Brillouin Scattering (SBS) Lasers},} in
  \emph{2018 IEEE Photonics Conference (IPC),}  (IEEE, 2018).

\end{thebibliography}

\end{document}


\title{Supplementary Material for\\``Simultaneous ponderomotive squeezing of light by two mechanical modes in an optomechanical system''}

\author{Peyman Malekzadeh\authormark{1}, Emil Zeuthen\authormark{1},Eric Langman\authormark{2}, Albert Schliesser\authormark{2}, Eugene Polzik\authormark{1,*} }

\address{\authormark{1}Niels Bohr Institute, University of Copenhagen, Blegdamsvej 17, 2100 Copenhagen, Denmark}
\address{\authormark{2}Center for Hybrid Quantum Networks, Niels Bohr Institute, University of Copenhagen, Blegdamsvej 17, Copenhagen 2100, Denmark}
\email{\authormark{*}polzik@nbi.ku.dk} 


Here we present the theoretical derivation of squeezing spectra, as well as squeezing plots for the second mechanical mode acquired for various laser drive detunings.
\section{Theoretical formulas for squeezing spectra}
In this section, we derive the spectral function fitted to the experimental data, acquired by direct detection, as presented in Fig.~2 in the main text; we also derive the optimal squeezing spectrum, representing the (greater) squeezing that could have been observed in a homodyne detection setup.

\begin{figure}[ht]
    \centering
    \includegraphics[width=1\linewidth]{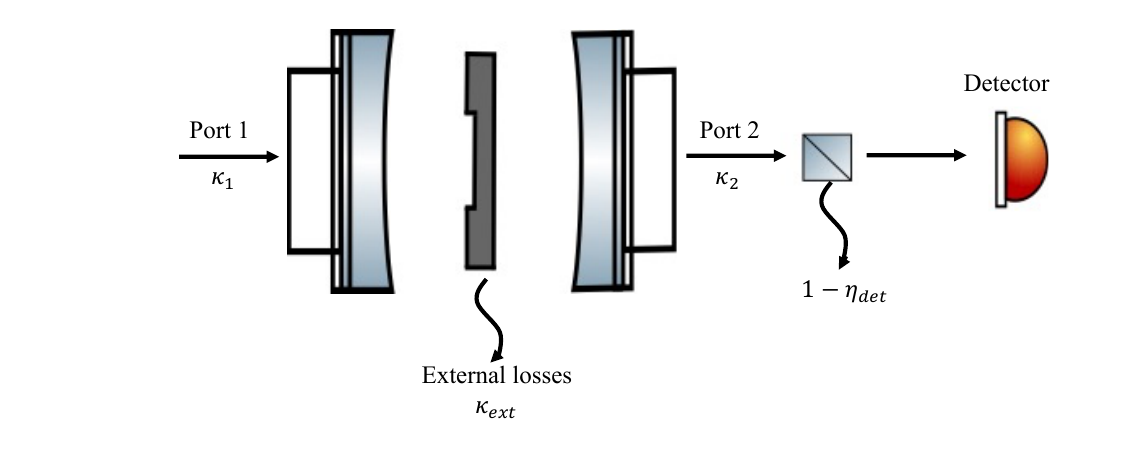}
    \caption{The cavity has three ports. Port 1 is the highly reflective mirror where light is coupled into the cavity. Port 2 is the highly transmissive mirror. The external cavity losses are specified by $\kappa_\mathrm{ext}$ and amounts to a third port. The detection loss is $1-\eta_\mathrm{det}$ and is modeled as a beam splitter in front of a perfect detector.}
    \label{fig:S1}
\end{figure}

The mathematical model is developed for a mechanical oscillator with two mechanical modes at frequencies $\Omega_{\mathrm{m}l}$, $l=1,2$; however, we focus on the limit relevant to the present experiment, where these two resonances are so far apart that no interference or hybridization between them arises. Naturally, each mechanical mode $l$ is described by a separate set of parameters including the optomechanical coupling rate $g_l$, intrinsic damping rate $\Gamma_{\mathrm{m}l}$, and thermal bath occupancy $n_{{\mathrm{th},l}}$. However, since the two mechanical modes couple to one and the same cavity mode, there is only one set of optical parameters, including the laser-drive detuning $\Delta=\omega_\mathrm{cav}-\omega_\mathrm{laser}$ and the cavity linewidth $\kappa = \sum_j \kappa_j$ in terms of the decay rates $\kappa_{j}$ corresponding to port $j=1, 2$, ext (see SM Fig.~\ref{fig:S1}).

We commence our derivation by formulating the equations of motion using Quantum Langevin equations in the Heisenberg picture. In the equations below, $\hat{Q}_l$ and $\hat{P}_l$ represent the position and momentum quadratures of mechanical mode $l$, $\hat{X}_{\mathrm{in},j}$ and $\hat{Y}_{\mathrm{in},j}$ represent the fluctuations on the amplitude and phase quadratures of light, while $\hat{P}_{\mathrm{in},l}$ is the thermal bath fluctuations impinging on the membrane. The equations of motion for the fluctuations of the cavity and mechanical modes are 
\begin{subequations}\label{eq-SM:EOMs}
\begin{align}
\dot{\hat{X}}&=-\frac{\kappa}{2}\hat{X}+\sum_{j}\sqrt{\kappa_{j}}\hat{X}_{\mathrm{in},j}+\Delta\hat{Y}\\
\dot{\hat{Y}}&=-\frac{\kappa}{2}\hat{Y}+\sum_{j}\sqrt{\kappa_{j}}\hat{Y}_{\mathrm{in},j}-\Delta\hat{X}-2\sum_l g_l\hat{Q}_l\\
\dot{\hat{Q}}_l&=\Omega_{\mathrm{m}l}\hat{P}_l\\
\dot{\hat{P}}_l&=-\Omega_{\mathrm{m}l}\hat{Q}_l-\Gamma_{\mathrm{m}l}\hat{P}_l+\sqrt{2\Gamma_{\mathrm{m}l}}\hat{P}_{\mathrm{in},l}-2g_l\hat{X}
\end{align}
\end{subequations}

By performing Fourier transformation ($\hat{O}(\omega)=\int_{-\infty}^{+\infty}\hat{O}(t)e^{i\omega{t}}dt$) of Eqs.~\eqref{eq-SM:EOMs}, it is straightforward to solve them algebraically. Solving Eqs.~(\ref{eq-SM:EOMs}a) and (\ref{eq-SM:EOMs}b) for the intracavity amplitude quadrature $\hat{X}(\omega)$ we find
\begin{equation}\label{eq-SM:X-Q}
\hat{X}(\omega) = \frac{1}{\left(\kappa/2 - i\omega\right)^2 + \Delta^2} \left[\sum_{j}\sqrt{\kappa_{j}}(\kappa/2 - i\omega)\hat{X}_{\mathrm{in},j}(\omega) + \Delta\sum_{j}\sqrt{\kappa_{j}}\hat{Y}_{\mathrm{in},j}(\omega) - 2\Delta \sum_l g_l\hat{Q}_l(\omega)\right].
\end{equation}
and for the mechanical position quadrature of mode $l$,  
\begin{multline}\label{eq-SM:Q}
\hat{Q}_l(\omega) = \chi_{\mathrm{eff}l}(\omega)\sqrt{2\Gamma_{\mathrm{m}l}}\hat{P}_{\mathrm{in},l}(\omega) - \frac{2g_l\chi_{\mathrm{eff}l}(\omega)}{\left(\kappa/2 - i\omega\right)^2 + \Delta^2}\\
 \times \left(\sum_{j}\sqrt{\kappa_{j}}\left[(\kappa/2 - i\omega)\hat{X}_{\mathrm{in},j}(\omega) + \Delta\hat{Y}_{\mathrm{in},j}(\omega)\right] - 2\Delta g_{l'\neq l}\hat{Q}_{l' \neq l}(\omega)\right),
\end{multline}
where $l'\neq l$, for given mechanical mode index $l=1,2$, is the index of the other mode; we have introduced the bare and effective mechanical susceptibilities 
\begin{equation}\label{eq-SM:chi-defs}
\chi_{\mathrm{m}l}(\omega)\equiv\frac{\Omega_{\mathrm{m}l}}{\Omega_{\mathrm{m}l}^2-\omega^2-i\Gamma_{\mathrm{m}l}\omega},\quad \chi_{\mathrm{eff}l}(\omega)\equiv\left({\chi_{\mathrm{m}l}^{-1}(\omega)
-\frac{4g_l^2\Delta}{(\kappa/2-i\omega)^2+\Delta^2}}\right)^{-1}.
\end{equation}
The term $\propto \hat{Q}_{l'\neq l}$ in Eq.~\eqref{eq-SM:Q} represents the coupling between mechanical modes 1 and 2 via the common cavity mode with which they interact; this can lead to mechanical mode hybridization. However, if the overlap of their effective susceptibilities is sufficiently small $\chi_{\mathrm{eff}1}(\omega)\chi_{\mathrm{eff}2}(\omega)\approx 0$ and $\sqrt{g_1 g_2} \ll |\Omega_{\mathrm{m},1}-\Omega_{\mathrm{m},2}|$, this effect is negligible. Assuming this to be the case, we will henceforth neglect the term $\propto \hat{Q}_{l'\neq l}$ in Eq.~\eqref{eq-SM:Q} and, for notational simplicity, confine attention to Fourier components near a particular mechanical mode $l$, $\omega\sim \Omega_{\mathrm{m}l}$.

\subsection{Direct detection spectrum $S_{XX}(\omega)$}\label{sec-SM:dir-det-spectrum}

The intracavity drive field can be represented by a classical displacement $\bar{X}$ of the amplitude quadrature fluctuations $\hat{X}\rightarrow \bar{X}+\hat{X}$ (in Eqs.~\eqref{eq-SM:EOMs} only its enhancement of the optomechanical coupling $g_l\propto\bar{X}$ is present). The transmitted drive field (i.e., leaving port 2) can be found from the input-output relation
\begin{equation}\label{eq-SM:in-out}
\hat{X}_{\mathrm{out},j}=\sqrt{\kappa_{j}}\hat{X}-\hat{X}_{\mathrm{in},j} 
\end{equation}
with $j=2$ to be simply $\bar{X}_{\mathrm{out},2} = \sqrt{\kappa_2}\bar{X}$. The measured photocurrent on the detector in terms of the annihilation operator of the field $\hat{a}_{\mathrm{out},2}=(\bar{X}_{\mathrm{out},2}+\hat{X}_{\mathrm{out},2}+i\hat{Y}_{\mathrm{out},2})/\sqrt{2}$ is determined by  
\begin{equation}\label{eq-SM:I-direct}
\hat{I}=\hat{a}_{\mathrm{out},2}^{\dagger}\hat{a}_{\mathrm{out},2}\approx \bar{X}_{\mathrm{out},2}^2/2 + \bar{X}_{\mathrm{out},2}\hat{X}_{\mathrm{out},2},
\end{equation}
where we have neglected terms quadratic in the (small) fluctuations $\hat{X}_{\mathrm{out},2}$ and $\hat{Y}_{\mathrm{out},2}$. Equation~\eqref{eq-SM:I-direct} shows that our direction-detection scheme is analogous to homodyne detection of the quadrature $\hat{X}_{\mathrm{out},2}$, and hence we will derive the associated spectrum $S_{X_{\mathrm{out},2}}$ below in order to compare with our experimentally recorded spectra.

To this end, we now substitute the expression for the mechanical position quadrature $\hat{Q}_l(\omega)$, Eq.~\eqref{eq-SM:Q}, into that of the light amplitude quadrature $\hat{X}(\omega)$, Eq.~\eqref{eq-SM:X-Q}, yields
\begin{multline}
    \hat{X}(\omega\sim \Omega_{\mathrm{m}l}) = \frac{-2g_l\Delta\chi_{\mathrm{eff}l}(\omega)\sqrt{2\Gamma_{\mathrm{m}l}}}{\left(\kappa/2 - i\omega\right)^2 + \Delta^2} \hat{P}_{\mathrm{in},l}(\omega) \\
    + \frac{\chi_{\mathrm{eff}l}(\omega)/\chi_{\mathrm{m}}(\omega)}{\left(\kappa/2 - i\omega\right)^2 + \Delta^2} 
    \sum_{j}\sqrt{\kappa_{j}}\left[\left(\kappa/2 - i\omega\right)\hat{X}_{\mathrm{in},j}(\omega) 
    + \Delta\hat{Y}_{\mathrm{in},j}(\omega)\right],
\end{multline}
having made use of the relationship $\chi_{\mathrm{eff}l}(\omega)/\chi_{\mathrm{m}l}(\omega)=1+4g_l^2\Delta\chi_{\mathrm{eff}l}(\omega)/[\left(\kappa/2 - i\omega\right)^2 + \Delta^2]$, cf.\ Eqs.~\eqref{eq-SM:chi-defs}.
Subsequently, we utilize the input-output relation~\eqref{eq-SM:in-out}  
to calculate the outgoing light amplitude quadrature for the output port (`2'),
\begin{equation}\label{eq-SM:X_out-sol}
\begin{aligned}
    \hat{X}_{\mathrm{out},2}(\omega\sim \Omega_{\mathrm{m}l}) = {}&\frac{-2g_l\Delta\chi_{\mathrm{eff}l}(\omega)\sqrt{2\Gamma_{\mathrm{m}l}}\sqrt{\kappa_{2}}}{\left(\kappa/2 - i\omega\right)^2 + \Delta^2} \hat{P}_{\mathrm{in},l}(\omega) \\
    &+ \frac{\sqrt{\kappa_{2}}\chi_{\mathrm{eff}l}(\omega)/\chi_{\mathrm{m}l}(\omega)}{\left(\kappa/2 - i\omega\right)^2 + \Delta^2} 
     \bigg[\left(\kappa/2 - i\omega\right) 
    \sum_{j\neq 2} \sqrt{\kappa_j}\hat{X}_{\mathrm{in},j}(\omega) 
    + \Delta 
    \sum_{j} \sqrt{\kappa_j}\hat{Y}_{\mathrm{in},j}(\omega)\bigg] \\
    &- \left[\frac{\left(\kappa/2 - i\omega\right)\kappa_{2}}{\left(\kappa/2 - i\omega\right)^2 + \Delta^2} \frac{\chi_{\mathrm{eff}l}(\omega)}{\chi_{\mathrm{m}l}(\omega)}
    - 1\right] \hat{X}_{\mathrm{in},2}(\omega).
\end{aligned}
\end{equation}

We use the symmetrized spectral density for an operator in the Fourier domain $\hat{O}(\omega)$,
\begin{equation}\label{eq-SM:sym-spectrum}
S[\hat{O}(\omega)] \equiv S_{O}(\omega)=(1/2)\int_{-\infty}^{+\infty}d\omega'\langle\hat{O}^{\dagger}(\omega)\hat{O}(\omega')+\hat{O}(\omega')\hat{O}^{\dagger}(\omega)\rangle,
\end{equation}
to calculate the power spectral density (PSD) of the amplitude quadrature of light at the output of the cavity. By using the same convention, we derive the input light noise PSD, which is assumed to be in their vacuum state, as $S_{X_{\mathrm{in},j}}=S_{Y_{\mathrm{in},j}}=1/2$, and for the PSD of the membrane's thermal fluctuations coupled to a thermal reservoir with occupancy of $n_\mathrm{th}$, we have $S_{P_{\mathrm{in},l}}=n_{\mathrm{th},l}+1/2$, we derive the direct detection spectrum,
\begin{multline}\label{eq-SM:S_X_out2}
    S_{X_{\mathrm{out},2}}(\omega\sim \Omega_{\mathrm{m}l}) = 
    \eta_{\mathrm{cav}}\left|\sqrt{\kappa}\frac{2\Delta g_l\chi_{\mathrm{eff}l}(\omega)\sqrt{2\Gamma_{\mathrm{m}l}}}{\left(\kappa/2 - i\omega\right)^2 + \Delta^2}\right|^2 
    \left(n_{\mathrm{th},l} + \frac{1}{2}\right) \\
    +\eta_{\mathrm{cav}} \left|\frac{\kappa\chi_{\mathrm{eff}l}(\omega)/\chi_{\mathrm{m}l}(\omega)}{\left(\kappa/2 - i\omega\right)^2 + \Delta^2} 
    \right|^2 
    \left[\frac{\kappa^2}{4} + \omega^2+ \Delta^2 \right]\frac{1}{2} 
    -\eta_{\mathrm{cav}}\text{Re}\left(\frac{(\kappa/2 - i\omega)\kappa}{\left(\kappa/2 - i\omega\right)^2 + \Delta^2}\frac{\chi_{\mathrm{eff}l}(\omega)}{\chi_{\mathrm{m}l}(\omega)}\right)
    + \frac{1}{2},
\end{multline}
as presented in the main text as Eq.~(5).

To develop the final model fitted to the experimental data, we adjust the aforementioned spectrum to account for losses and compute the detected spectrum
\begin{equation}\label{eq-SM:S-det}
    S_\mathrm{det}(\omega)=\eta_\mathrm{det}S_{X_{\mathrm{out},2}}(\omega)+(1 - \eta_\mathrm{det})\frac{1}{2}.
\end{equation}
Note that this amounts to replacing $\eta_{\mathrm{cav}}\rightarrow \eta_{\mathrm{net}}\equiv\eta_{\mathrm{det}}\eta_{\mathrm{cav}}$ in Eq.~\eqref{eq-SM:S_X_out2}, i.e., cavity and detection losses are equivalent in the manner in which they degrade the observed squeezing.

\subsection{Spectrum of optimal squeezing}

As discussed in Sec.~3 of the main text, a full characterization of quadrature squeezing involves considering a general (rotated) output quadrature of the form
\begin{equation}\label{eq-SM:X-theta}
\hat{X}_{\mathrm{out,2}}^{(\theta)}(\omega)\equiv\hat{X}_{\mathrm{out,2}}(\omega)\cos\theta+\hat{Y}_{\mathrm{out,2}}(\omega)\sin\theta. 
\end{equation}
To determine the quadrature phase $\theta_{\mathrm{min}}(\omega)$ that minimizes $S[\hat{X}_{\mathrm{out,2}}^{(\theta)}(\omega)]$ for given $\omega$, i.e., shows maximal squeezing, we must consider $S_{X_{\mathrm{out},2}}(\omega)$ [Eq.~\eqref{eq-SM:S_X_out2}] together with $S_{Y_{\mathrm{out},2}}(\omega)$ and the (complex-valued) symmetrized cross-spectral density $S_{X_{\mathrm{out},2}Y_{\mathrm{out},2}}(\omega)$, defined for two Fourier domain operators $\hat{O}_1$ and $\hat{O}_2$ as
\begin{equation}\label{eq-SM:sym-cross-spectrum}
S_{O_1 O_2}(\omega)=(1/2)\int_{-\infty}^{+\infty}d\omega'\langle\hat{O}_1^{\dagger}(\omega)\hat{O}_2(\omega')+\hat{O}_2(\omega')\hat{O}_1^{\dagger}(\omega)\rangle.
\end{equation}
Evaluating the spectrum of $\hat{X}_{\mathrm{out,2}}^{(\theta)}(\omega)$ [Eq.~\eqref{eq-SM:X-theta}] using the definition~\eqref{eq-SM:sym-spectrum} we find 
\begin{multline}\label{eq-SM:S_X-out-theta}
S_{X_{\mathrm{out},2}^{(\theta)}}(\omega) =\cos^{2}(\theta)S_{X_{\mathrm{out},2}}(\omega)+\sin^{2}(\theta)S_{Y_{\mathrm{out},2}}(\omega)+2\text{sin}(\theta)\text{cos}(\theta)\text{Re}[S_{X_{\mathrm{out},2}Y_{\mathrm{out},2}}(\omega)] \\
	=\frac{1}{2}\left(S_{X_{\mathrm{out},2}}(\omega)+S_{Y_{\mathrm{out},2}}(\omega)\right)+\cos(2\theta)\frac{1}{2}\left(S_{X_{\mathrm{out},2}}(\omega)-S_{Y_{\mathrm{out},2}}(\omega)\right)+\text{sin}(2\theta)\text{Re}[S_{X_{\mathrm{out},2}Y_{\mathrm{out},2}}(\omega)].
\end{multline}
We note the familiar characteristic of homodyne detection that only the real (in-phase) correlations between the quadratures $\text{Re}[S_{X_{\mathrm{out},2}Y_{\mathrm{out},2}}(\omega)]$ contribute to the observed squeezing.
From Eq.~\eqref{eq-SM:S_X-out-theta}, the optimal quadrature phase $\theta_{\mathrm{min}}(\omega)$, i.e., exhibiting maximal squeezing for given $\omega$, is seen to obey the set of equations
\begin{equation}
    \tan[2\theta_{\mathrm{min}}(\omega)] = \frac{2\text{Re}[S_{X_{\mathrm{out},2}Y_{\mathrm{out},2}}(\omega)]}{S_{X_{\mathrm{out},2}}(\omega)-S_{Y_{\mathrm{out},2}}(\omega)}\enspace\wedge\enspace \text{sign}\{\sin[2\theta_{\mathrm{min}}(\omega)]\} = -\text{sign}\{\text{Re}[S_{X_{\mathrm{out},2}Y_{\mathrm{out},2}}(\omega)]\}.
\end{equation}
The associated spectrum of optimal squeezing results from setting $\theta=\theta_{\mathrm{min}}(\omega)$ in Eq.~\eqref{eq-SM:S_X-out-theta}  resulting in
\begin{multline}\label{eq-SM:opt-squeezing-spectrum}
    \left.S_{X_{\mathrm{out},2}^{(\theta)}}(\omega)\right|_{\theta=\theta_{\mathrm{min}}(\omega)} = \\ \frac{1}{2}\left(S_{X_{\mathrm{out},2}}(\omega)+S_{Y_{\mathrm{out},2}}(\omega)\right) - \sqrt{\frac{1}{4}\left(S_{X_{\mathrm{out},2}}(\omega)-S_{Y_{\mathrm{out},2}}(\omega)\right)^2+\text{Re}^2[S_{X_{\mathrm{out},2}Y_{\mathrm{out},2}}(\omega)]}.
\end{multline}
This type of spectrum could be constructed experimentally by, e.g., for given $\omega$, setting the homodyne detection phase equal to $\theta_{\mathrm{min}}(\omega)$, and in this manner acquiring the spectral amplitude for each $\omega$ separately. 
The theoretical predictions for the degree of squeezing that could have been observed in homodyne detection are found by numerically minimizing Eq.~\eqref{eq-SM:opt-squeezing-spectrum} over $\omega$. Note that the inclusion of the finite detection efficiency $\eta_{\mathrm{det}}$ does not affect $\theta_{\mathrm{min}}(\omega)$, and, in fact, Eq.~\eqref{eq-SM:S-det} can simply be applied to Eq.~\eqref{eq-SM:opt-squeezing-spectrum} to incorporate this effect in our calculation.

In order to apply Eq.~\eqref{eq-SM:opt-squeezing-spectrum}, we derive formulas for $S_{Y_{\mathrm{out},2}}(\omega)$ and $\text{Re}[S_{X_{\mathrm{out},2}Y_{\mathrm{out},2}}(\omega)]$ following steps similar to those in SM Sec.~\ref{sec-SM:dir-det-spectrum}.
Using the input-output relation describing the phase quadrature of the output port (`2')
\begin{equation}\label{eq-SM:in-out-Y}
\hat{Y}_{\mathrm{out},2}=\sqrt{\kappa_{2}}\hat{Y}-\hat{Y}_{\mathrm{in},2}. 
\end{equation}
we can arrive at
\begin{multline}\label{eq-SM:Y_out-sol}
\hat{Y}_{\mathrm{out},2}(\omega\sim\Omega_{\mathrm{m},l})  =-\sqrt{\kappa_{2}}\frac{2(\kappa/2-i\omega)g_l\chi_{\mathrm{eff}l}(\omega)\sqrt{2\Gamma_{\mathrm{m}l}}}{(\kappa/2-i\omega)^{2}+\Delta^{2}}\hat{P}_{\mathrm{in},l}(\omega)\\
+\sqrt{\kappa_{2}}\frac{\chi_{\mathrm{eff}l}(\omega)/\chi_{\mathrm{m}l}(\omega)}{(\kappa/2-i\omega)^{2}+\Delta^{2}}\left((\kappa/2-i\omega)\sum_{j\neq2}\sqrt{\kappa_{j}}\hat{Y}_{\mathrm{in},j}(\omega)+\left(4g_l^{2}\chi_{\mathrm{m}l}(\omega)-\Delta\right)\sum_{j}\sqrt{\kappa_{j}}\hat{X}_{\mathrm{in},j}(\omega)\right)\\
+\left[\frac{(\kappa/2-i\omega)\kappa_{2}}{(\kappa/2-i\omega)^{2}+\Delta^{2}}\frac{\chi_{\mathrm{eff}l}(\omega)}{\chi_{\mathrm{m}l}(\omega)}-1\right]\hat{Y}_{\mathrm{in},2}(\omega)
\end{multline}
and using Eq.~\eqref{eq-SM:sym-spectrum} we arrive at the spectrum for the phase quadrature of the output port
\begin{equation}\label{eq-SM:S_Y_out2}
\begin{aligned}
S_{Y_{\text{out},2}}(\omega\sim\Omega_{\mathrm{m},l}) = {}&\eta_{\mathrm{cav}}\left|\sqrt{\kappa}\frac{2(\kappa/2-i\omega)g_l\chi_{\mathrm{eff}l}(\omega)\sqrt{2\Gamma_{\mathrm{m}l}}}{(\kappa/2-i\omega)^{2}+\Delta^{2}}\right|^{2}\left(n_{\mathrm{th},l}+\frac{1}{2}\right)\\
 &+\eta_{\mathrm{cav}}\left|\frac{\kappa\chi_{\mathrm{eff}l}(\omega)/\chi_{\mathrm{m}l}(\omega)}{(\kappa/2-i\omega)^{2}+\Delta^{2}}\right|^{2}\left(\frac{\kappa^{2}}{4}+\omega^{2}+\left|4g_l^{2}\chi_{\mathrm{m}l}(\omega)-\Delta\right|^{2}\right)\frac{1}{2}\\
 &-\eta_{\mathrm{cav}}\text{Re}\left(\frac{(\kappa/2-i\omega)\kappa}{(\kappa/2-i\omega)^{2}+\Delta^{2}}\frac{\chi_{\mathrm{eff}l}(\omega)}{\chi_{\mathrm{m}l}(\omega)}\right)+\frac{1}{2}.
\end{aligned}
\end{equation}

Finally, substituting Eqs.~\eqref{eq-SM:X_out-sol} and \eqref{eq-SM:Y_out-sol} into Eq.~\eqref{eq-SM:sym-cross-spectrum}, we find the in-phase cross-correlation of the $\hat{X}_{\mathrm{out},2}$ and $\hat{Y}_{\mathrm{out},2}$ quadratures 
\begin{equation}\label{eq-SM:S_XY_out2}
\begin{aligned}
\text{Re}[S_{X_{\mathrm{out},2}Y_{\mathrm{out},2}}(\omega)] ={}&\eta_{\mathrm{cav}}\left|\kappa\frac{2g_l\chi_{\mathrm{eff}l}(\omega)}{\left(\kappa/2-i\omega\right)^{2}+\Delta^{2}}\right|^{2}\left(\frac{\Delta}{2}2\Gamma_{\mathrm{m}l}\left(n_{\mathrm{th},l}+\frac{1}{2}\right)+\text{Re}\left[\left(\kappa/2-i\omega\right)\chi_{\mathrm{m}l}^{-1}(\omega)\right]\frac{1}{2}\right)\\
&-\eta_{\mathrm{cav}}\text{Re}\left[\kappa\frac{4g_l^{2}\chi_{\mathrm{eff}l}(\omega)}{(\kappa/2-i\omega)^{2}+\Delta^{2}}\right]\frac{1}{2}.  
\end{aligned}
\end{equation}
This completes the derivation of the functions \eqref{eq-SM:S_X_out2}, \eqref{eq-SM:S_Y_out2}, and \eqref{eq-SM:S_XY_out2} required to evaluate Eq.~\eqref{eq-SM:opt-squeezing-spectrum}.

\newpage
\section{Squeezing spectra corresponding to the second band gap for various detuning values}
\begin{figure}[ht]
    \centering
    \includegraphics[width=1\linewidth]{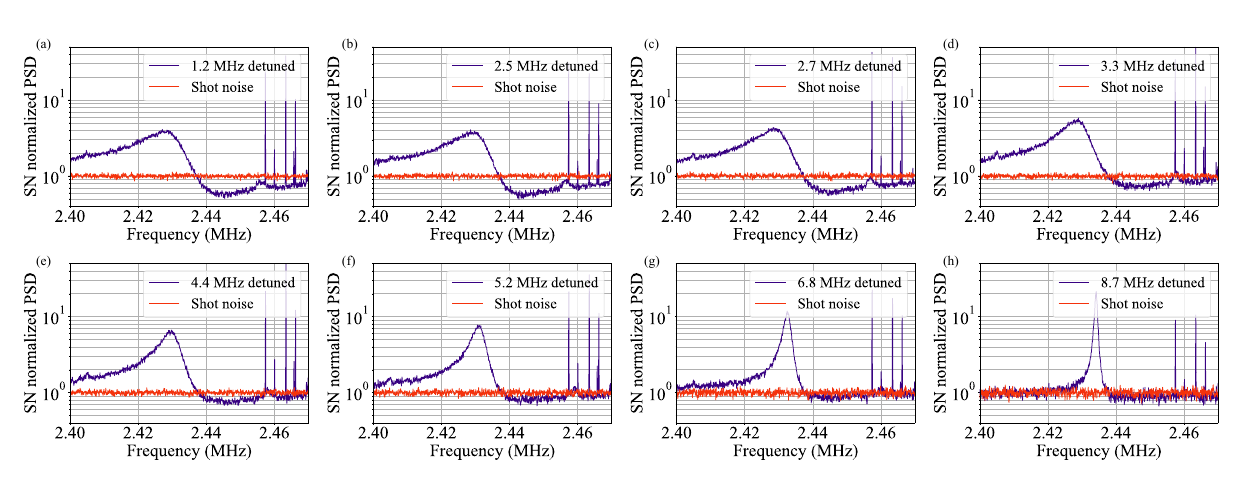}
    \caption{Cavity output spectra near the second mechanical mode for different detuning ranging from 1.2\,MHz to 8.7\,MHz.}
    \label{fig:S2}
\end{figure}

SM Figure~\ref{fig:S2} presents the squeezing spectra corresponding to the second band gap for various detuning values, following the approach outlined in Fig.~3 of the main text. As described in Sec.~4 of the main text, the cavity was locked to different detunings ranging from 1.2\,MHz to 8.7\,MHz for the 2.43\,MHz mechanical mode, while the input power coupled to the cavity was maintained at 3\,mW. The detuning for each measurement was determined by phase modulating the input light using an electro-optic modulator. 
